# A Brief Review of Continuous Models for Ionic Solutions: the Poisson-Boltzmann and Related Theories

Mao Su（苏茂）[1,2] and Yanting Wang（王延颋）[1,2]*

[1]CAS Key Laboratory of Theoretical Physics, Institute of Theoretical Physics, Chinese Academy of Sciences, 55 East Zhongguancun Road, P. O. Box 2735, Beijing 100190, China

[2]School of Physical Sciences, University of Chinese Academy of Sciences, 19A Yuquan Road, Beijing 100049, China

**Abstract:** The Poisson-Boltzmann (PB) theory is one of the most important theoretical models describing charged systems continuously. However, it suffers from neglecting ion correlations, which hinders its applicability to more general charged systems other than extremely dilute ones. Therefore, some modified versions of the PB theory are developed to effectively include ion correlations. Focused on their applications to ionic solutions, the original PB theory and its variances, including the field-theoretic approach, the correlation-enhanced PB model, the Outhwaite-Bhuiyan modified PB theory and the mean field theories, are briefly reviewed in this paper with the diagnosis of their advantages and limitations.

## 1. Introduction

Theoretical descriptions of charged systems are very important for a wide range of applications, such as the interfacial phenomena of aqueous solutions [1-3] and the structure and dynamics of biomolecules [4-6]. Among various kinds of theories, the Poisson-Boltzmann (PB) theory initially developed by Debye and Hückel [7] in 1923 is the most famous one. The PB theory is particularly useful in depicting ionic solutions, for which it serves as a mesoscopic continuous model describing the average potential distribution of ionic solutions. Once the potential distribution is determined, physical properties such as activity and osmotic coefficient can be calculated with the knowledge of statistical physics. Because of their simplicity, the PB theory and its linearized version, the Debye-Hückel (DH) theory, are widely used in many scientific occasions, including investigating the electrostatic and thermodynamic properties, such as activity coefficient, solvation energy, and so on [8,9], of ionic solutions and charged molecules.

It is well known that the PB theory is on the basis of key approximations we will introduce below that limit its applicability to extremely dilute solutions whose ion correlation effect is negligible. Therefore, the PB model can completely fail in some practical scenarios [10]. For example, DNA and many protein molecules are so highly charged that the electrostatic correlation effect cannot be neglected [6,11]. It is worth noting that the PB theory may in some occasions, e.g., calculating thermodynamic properties for densely charged systems, provide reasonable results due to error cancellation. However, it is by chance and cannot serve as a routine treatment to dense systems.

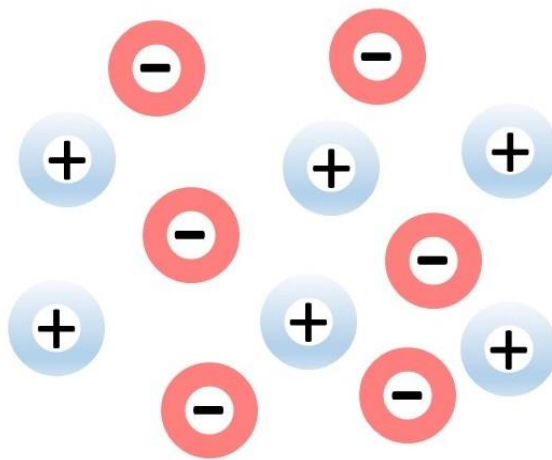

**Figure 1.** Schematic of a charged system which can be described by the PB theory. When the correlation effect illustrated by the halos around point charges is significant, the original PB theory may fail and enhanced PB theories are on demand.

As illustrated in Figure 1, in order to yield more accurate solutions for a wider range of charged systems, a variety of theoretical models, either based on the PB theory or not, have been developed [12-14]. Interestingly, most of them are connected to the PB theory, either through the form of equation or through the form of solution, which in turn manifests the importance of the PB theory.

This review is organized as follows. The derivation of the original PB theory is shown in section 2, along with detailed discussion of the approximations used in the derivation. In section 3, the field-theoretic (FT) approach for calculating the partition function is reviewed, and as a consequence a set of self-consistent equations are obtained by the variational method. In section 4, we introduce our correlation-enhanced PB model along with the Outhwaite-Bhuiyan modified PB (MPB) theories. Finally, a set of mean field theories including the dressed-ion theory (DIT) and the molecular Debye-Hückel (MDH) theory are reviewed in section 5. In the Appendix, the integral equation theory is briefly introduced as the concepts are used in the MPB theories as well as the mean field theories. The SI unit is used by default unless the use of the Gaussian unit is explicitly declared. It should be noted that this brief review is more titled to serve as a manual for beginners, so some most recent advanced theoretical and numerical progresses, such as the concentration-dependent dielectric model [15,16], are not reviewed due to the limited space.

**2. The Poisson-Boltzmann (PB) theory**

The PB theory is originally derived by Debye and Hückel in 1923 [7]. It is formally a Poisson's equation with the charge density distribution obeying the Boltzmann distribution. The original PB equation is a non-linear equation which is usually hard to solve analytically. Debye and Hückel then implement some approximations to linearize the original PB equation, leading to the linearized PB equation, or the DH equation [17], which can be solved analytically and end up with the Yukawa potential. The activity coefficients calculated by the DH equation at the low-concentration limit fit the experimental values very well, but deviate more severely at a higher concentration due to the approximations incorporated in the DH equation [17].

**2.1. Derivation of the original PB theory**

The derivation of the PB theory can be found in Ref. [17] and we summarize it as follows. The first approximation made in the PB theory is neglecting the inner structure of particles. For an aqueous ionic solution, water is treated as a continuous background with the relative dielectric constant $\varepsilon$, and ions are assumed to be point charges. For an infinitely large domain consisted of $N$ ions located at $r_1, r_2, \cdots, r_N$, the electric potential of the system at $r$ is

$$\psi(r) = \sum_{i=1}^{N} \frac{q_i}{4\pi\varepsilon_0\varepsilon|r - r_i|}, \tag{1}$$

which satisfies Poisson's equation

$$-\nabla^2 \psi(r) = \frac{1}{\varepsilon_0\varepsilon} \rho(r), \tag{2}$$

where $\varepsilon_0$ is the dielectric constant of vacuum, $q_i$ is the charge of the $i$th ion, and $\rho(r)$ is the charge density at $r$. In a homogeneous charge-neutral system, the solution of $\psi(r)$ is zero everywhere. To obtain useful information, we fix an ion at $r_1$ and study the electrostatic potential at $r_2$:

$$\psi(r_2, r_1) = \frac{q_1}{4\pi\varepsilon_0\varepsilon|r_2 - r_1|} + \sum_{i=3}^{N} \frac{q_i}{4\pi\varepsilon_0\varepsilon|r_2 - r_i|}. \tag{3}$$

Since the positions of other ions fluctuate, we find the canonical ensemble average of $\psi(r_2, r_1)$ to be:

$$\langle \psi(r_2, r_1) \rangle = \frac{\int \cdots \int \psi(r_2, r_1) e^{-\beta U} dr_3 \cdots dr_N}{\int \cdots \int e^{-\beta U} dr_3 \cdots dr_N}, \tag{4}$$

where $U = U(r_1, r_2, ..., r_N)$ is the potential energy and $\beta \equiv \dfrac{1}{k_B T}$ with $k_B$ the Boltzmann constant and $T$ the temperature. Taking the Laplacian with respect to $r_2$ on both sides:

$$\nabla_{r_2}^2 \langle \psi(r_2, r_1) \rangle = \frac{\int \cdots \int \nabla_{r_2}^2 \psi(r_2, r_1) e^{-\beta U} dr_3 \cdots dr_N}{\int \cdots \int e^{-\beta U} dr_3 \cdots dr_N}$$

$$= \frac{\int \cdots \int -\dfrac{1}{\varepsilon_0 \varepsilon} \rho(r_2, r_1) e^{-\beta U} dr_3 \cdots dr_N}{\int \cdots \int e^{-\beta U} dr_3 \cdots dr_N} \qquad (5)$$

$$= -\frac{1}{\varepsilon_0 \varepsilon} \langle \rho(r_2, r_1) \rangle.$$

The average density distribution $\langle \rho(r_2, r_1) \rangle$ is related to the radial distribution function (RDF) $g(r_2, r_1)$ by

$$\langle \rho(r_2, r_1) \rangle = \sum_s c_s q_s g_{1s}(r_2, r_1), \qquad (6)$$

where $c_s$ is the bulk number concentration of the $s$-type ions and $g_{1s}(r_2, r_1)$ is the RDF of $s$-type ions about the central ion located at $r_1$. The potential of mean force (PMF) was utilized for finding the solution of the above equation [18]. We Consider the mean force exerted on the ion at $r_2$ with the fixed ion at $r_1$:

$$-\left\langle \frac{d}{dr_2} U \right\rangle = \frac{-\int \cdots \int \dfrac{dU}{dr_2} e^{-\beta U} dr_3 \cdots dr_N}{\int \cdots \int e^{-\beta U} dr_3 \cdots dr_N}$$

$$= \frac{1}{\beta} \frac{\int \cdots \int \dfrac{d}{dr_2} e^{-\beta U} dr_3 \cdots dr_N}{\int \cdots \int e^{-\beta U} dr_3 \cdots dr_N}$$

$$= \frac{1}{\beta} \frac{d}{dr_2} \ln \left( \int \cdots \int e^{-\beta U} dr_3 \cdots dr_N \right) \qquad (7)$$

$$= \frac{1}{\beta} \frac{d}{dr_2} \ln \left( \frac{N(N-1) \int \cdots \int e^{-\beta U} dr_3 \cdots dr_N}{Z} \right)$$

$$\equiv \frac{1}{\beta} \frac{d}{dr_2} \ln g(r_2, r_1).$$

In the above derivation, the ion number $N$ and the partition function $Z = \int \cdots \int e^{-\beta U} dr_1 \cdots dr_N$ are independent of $r_2$, so $N(N-1)/Z$ can be inserted to obtain the definition of the RDF. We then define

$$w(r_2, r_1) \equiv -\frac{1}{\beta} \ln g(r_2, r_1). \tag{8}$$

Obviously the negative gradient of $w(r_2, r_1)$ is the mean force, so $w(r_2, r_1)$ is called the PMF. Combining Eqs. (5), (6), (8) together, we have

$$\nabla^2 \langle \psi(r_2, r_1) \rangle = -\frac{1}{\varepsilon_0 \varepsilon} \sum_s c_s q_s e^{-\beta w_{1s}(r_2, r_1)}. \tag{9}$$

Because the PMF $w(r_2, r_1)$ of a homogeneous system depends only upon $r \equiv |r_2 - r_1|$, for simplicity, we replace the denotation $(r_2, r_1)$ by $(r)$. The second approximation is that the PMF $w_{1s}(r)$ is replaced by $q_s \psi(r)$, where the mean potential energy $\psi(r) \equiv \langle \psi(r_2, r_1) \rangle$, which leads to the PB equation:

$$\nabla^2 \psi(r) = -\frac{1}{\varepsilon_0 \varepsilon} \sum_s c_s q_s e^{-\beta q_s \psi(r)}. \tag{10}$$

When the third approximation of linearization (expanding the exponent to the first order) is taken, the PB theory is further simplified to be the analytically solvable DH equation:

$$\nabla^2 \psi(r) = \frac{\beta}{\varepsilon_0 \varepsilon} \sum_s c_s q_s^2 \psi(r). \tag{11}$$

The solution of the DH equation for a point charge in an infinitely large domain is a screened Coulomb potential, named the Yukawa potential:

$$\psi(r) = \frac{q_1}{4\pi \varepsilon_0 \varepsilon r} e^{-\kappa r}, \tag{12}$$

where the Debye screening length $1/\kappa$ is defined by

$$\kappa^2 \equiv \frac{\beta}{\varepsilon_0 \varepsilon} \sum_s c_s q_s^2. \tag{13}$$

The physical properties such as RDF, activity coefficient, and osmotic coefficient can be calculated once the potential is known [17].

The PB equation is difficult to be solved due to the non-linearity. However, in some special cases it is possible to find the analytical solutions. For example, the PB theory can be applied to electrolytes with a charged plate, and the result is known as the Gouy-Chapman theory [19]. More generally, the PB equation is solved by numerical methods such as the finite difference method (FDM), finite element method (FEM), and so on. Accelerate techniques such as conjugate gradient method, fast multi-pole method, and fast Fourier transform (FFT)-based approaches are also used to solve the PB equation numerically [20].

**2.2. Limitations**

The PB theory neglects molecular structures (exclusive volume effects) of ions and solvents. As a result, the structural properties calculated by the PB theory may deviate significantly from the real solutions. For example, the RDF for an ionic aqueous solution described by the PB theory is essentially the density distribution with respect to a hard sphere and the central ion is treated as a boundary condition, so the RDF has the feature of a gas, i.e., with only one peak and decaying monotonically [21]. In real solutions, however, a typical RDF always has a valley at about 0.5 nm, which is about the size of a water molecule, because that space is occupied explicitly by water molecules [22].

In most application scenarios, however, the exclusive volume effect only has a minor influence because thermodynamic properties such as solvation energy are the major focus of interest. Implicit solvent models are studied by researchers to evaluate the solute PMF, which determines the statistical weight of solute conformations and therefore allows molecular simulations without explicit solvent molecules [23]. When the solution goes to the high concentration limit, i.e. near close packing, the PB theory completely fails due to the significant exclusive volume effect. In this case, the integral equation theory (see Appendix) with the Percus–Yevick (PY) or Hypernetted Chain (HNC) approximation can be employed to calculate the distribution function [24]. In particular, the HNC approximation is known to be very accurate for the charged hard-sphere model. However, the integral equation method cannot be applied to problems involving boundaries.

The second approximation of replacing the PMF by the potential energy $q\psi(r)$ in the PB theory neglects ion correlations, which is however most of the time important due to the long-range nature of electrostatic interaction. Many phenomena such as overcharge [25-29], like-charge attraction [30-34], and opposite-charge repulsion [35] have correlations and thus cannot be satisfactorily described by the PB theory. Particularly, the charge effect plays an essential role in most biomolecular systems, but many of them, such as protein and DNA [4-6,25], cannot be effectively described by the PB theory because they are highly charged with strong ion correlations.

**2.3. Asymmetry problem**

It should be noted that there is an inconsistency in the PB equation related with the approximation of $w(r) \approx q\psi(r)$, as firstly revealed by Onsager [36]. Let us consider the average concentration of ions of species $i$ at a distance $r$ from a given ion of species $j$:

$$n_{ji}(r) = n_i \exp(-\beta w_{ji}(r)), \tag{14}$$

and vice versa

$$n_{ij}(r) = n_j \exp(-\beta w_{ij}(r)), \tag{15}$$

where $n_i$ is the number density of species $i$ under no constraint. We expect that $w_{ij}(r) = w_{ji}(r)$ since $w$ is the reversible work expended in bringing the two ions $i$ and $j$ from infinity to distance $r$. The PB theory is on the basis of the approximation that

$$w_{ji}(r) = q_i \psi_j(r) \tag{16}$$

according to Eq. (14) and

$$w_{ij}(r) = q_j \psi_i(r) \tag{17}$$

according to Eq. (15). One may expect that

$$q_i \psi_j(r) = q_j \psi_i(r). \tag{18}$$

The above cannot be true except when $i = j$. Therefore, the PB solution cannot hold due to nonlinearity except in the symmetric case.

However, it is easy to show that the Yukawa potential in Eq. (12), the solution for the DH equation, satisfies Eq. (18). In fact, the DH equation is practically more frequently used because it is linear and has analytic solutions. The most successful application of the DH equation is the calculation of activity coefficient $\gamma_\pm$, which agrees very well with experiment at the low concentration limit [17],

$$\ln \gamma_\pm = -|q_+ q_-| \frac{\kappa}{8\pi\varepsilon_0 \varepsilon k_B T}. \tag{19}$$

Therefore, the DH theory is called the exact limiting law for ionic solutions. Although it is originated from error cancellation, this result suggests that the DH theory is a necessary step in deriving the exact limiting theory. Therefore, the Yukawa potential can be used to develop mean field descriptions for ionic solutions. The DIT [37,38] developed by Kjellander and Mitchell has an asymptotic solution taking the form of the Yukawa potential but including effective charges and a dielectric constant, which is exact as long as the effective parameters are correctly determined. The MDH theory [39,40] developed by Xiao and Song suggests a multi-Yukawa potential, which is also exact if the coefficients can be correctly obtained.

**3. Field-theoretic (FT) approach**

The FT approach provides a way for solving the many-body partition function by transforming the partition function into a functional integration [41-43] through the Hubbard-Stratonovich (HS) transformation [44]. The functional form of the partition function can then be solved by various methods. For example, the simplest approximation is known as the saddle-point approximation which recovers the original PB equation; it can also be solved with the perturbation method [45] and the variational methods [46-50] to obtain more accurate self-consistent equations for depicting ionic systems.

## 3.1 Formulism

The HS transformation is the key of the FT approach. To understand it, we consider a toy model with the grand canonical partition function:

$$\Xi = \sum_{N=0}^{\infty} \frac{\lambda_C^N}{N!} e^{-\varepsilon_C N^2/2}, \tag{20}$$

where $\lambda_C$ and $\varepsilon_C$ are just two mathematical parameters. In order to solve the partition function, one can transform the summation into integrals by noting the two identities

$$e^{-\varepsilon_C N^2/2} = \frac{1}{\sqrt{2\pi\varepsilon}} \int_{-\infty}^{\infty} dx\, e^{-x^2/2\varepsilon_C} e^{ixN}, \tag{21}$$

$$e^x = \sum_{N=0}^{\infty} \frac{x^N}{N!}. \tag{22}$$

The partition function can then be written as

$$\begin{aligned}\Xi &= \sum_{N=0}^{\infty} \frac{\lambda_C^N}{N!} e^{-\varepsilon_C N^2/2} \\ &= \frac{1}{\sqrt{2\pi\varepsilon_C}} \int_{-\infty}^{\infty} dx\, e^{-x^2/2\varepsilon_C} \sum_{N=0}^{\infty} \frac{\lambda_C^N}{N!} e^{ixN} \\ &= \frac{1}{\sqrt{2\pi\varepsilon_C}} \int_{-\infty}^{\infty} dx\, \exp\left(-\left(x^2/2\varepsilon_C - \lambda_C e^{ix}\right)\right).\end{aligned} \tag{23}$$

The Gaussian identity can be generalized to multiple species as

$$e^{-\frac{1}{2}\varepsilon_{Cij}N_iN_j} \equiv e^{-\frac{1}{2}(N^{\mathrm{T}}\varepsilon_C N)} = \int_{-\infty}^{\infty} \frac{\prod_{i=1}^{M} dx_i}{\sqrt{\det 2\pi\varepsilon_C}} e^{-\frac{1}{2}(x^{\mathrm{T}}\varepsilon_C^{-1}x)} e^{iN^{\mathrm{T}}x}, \tag{24}$$

and to the continuous case [41] as

$$e^{-\frac{1}{2}\int drdr'\rho(r)\varepsilon_C(r,r')\rho(r')} = \int \frac{D\xi}{\sqrt{\det 2\pi\varepsilon_C}} e^{-\frac{1}{2}\int drdr'\xi(r)\varepsilon_C^{-1}(r,r')\xi(r')} e^{i\int dr\rho(r)\xi(r)}, \tag{25}$$

where $\xi(r)$ is a field variable used to decouple the quadratic interaction on the left hand side.

Let us consider a typical charged particle system consisting of $n_+$ cations and $n_-$ anions whose Hamiltonian is

$$H = \frac{e^2}{2} \int drdr'\, \rho(r) C(r,r') \rho(r'), \tag{26}$$

where $e\rho(r)$ is the total charge density and $C(r,r')$ is the Coulomb operator defined by

$$-\nabla \cdot \left[\varepsilon_0 \varepsilon \nabla C(r,r')\right] = \delta(r,r'). \tag{27}$$

The grand partition function of a charged particle system is

$$\Xi = \sum_{n_+=0}^{\infty} \sum_{n_-=0}^{\infty} Q(n_+, n_-) e^{n_+\mu_+} e^{n_-\mu_-}, \tag{28}$$

where $n_\pm$ are the number densities of cations and anions, $\mu_\pm$ are the chemical potentials, and $Q(n_+, n_-)$ is the canonical partition function:

$$Q(n_+, n_-) = \frac{1}{n_+! n_-!} \int \prod_{i=1}^{n_+} dr_i \prod_{j=1}^{n_-} dr_j \exp(-\beta H). \tag{29}$$

Applying the HS transformation Eq. (25) and define a dimensionless field $\varphi \equiv \beta e \xi$, the grand partition function becomes [47]

$$\Xi = \frac{1}{Z_C} \int D\varphi e^{\{-L[\varphi]\}}, \tag{30}$$

where $Z_C$ is the associated partition function. The action $L[\varphi]$ in the functional integral is defined as

$$L[\varphi] = \int dr \left[\frac{1}{2}\epsilon \varphi(r)(-\nabla^2)\varphi(r) - \lambda_+ e^{-iz_+\varphi} - \lambda_- e^{-iz_-\varphi}\right], \tag{31}$$

where $\epsilon \equiv \varepsilon_0 \varepsilon / (\beta e^2)$ is the rescaled permittivity, $z_\pm$ are the valences of ions, and $\lambda_\pm$ are the fugacities [47]. The saddle point for the action, $\left.\frac{\delta L}{\delta \varphi}\right|_{\varphi=\varphi_0} = 0$, can be obtained by the Euler-Lagrange equation

$$-\epsilon \nabla^2 \varphi_0 = i\lambda_+ e^{-iz_+\varphi_0} + i\lambda_- e^{-iz_-\varphi_0}, \tag{32}$$

which retrogresses into the PB equation if $i\varphi_0$ is regarded as the average potential.

### 3.2. Approximate solutions

The grand partition function Eq. (30) can be solved with more sophisticated methods to obtain a modified PB equation including fluctuation and correlation effects. The variational methods are used by several groups to obtain modified PB equations. The most commonly used reference action has the Gaussian form [47,48,51]:

$$L_{\text{ref}}[\varphi] = \frac{1}{2}\int \mathrm{d}r\mathrm{d}r' [\varphi(r)+i\phi(r)]G^{-1}(r,r')[\varphi(r')+i\phi(r')], \tag{33}$$

where $\varphi(r)$ is the dimensionless field appeared in Eq. (30), $G(r,r')$ is a correlation function which could be the incremental potential in section 4.1, and $\phi(r) \equiv \beta e\psi(r)$ corresponds to the average electrostatic field. The variational free energy is

$$F = F_{\text{ref}} + \langle L[\varphi] - L_{\text{ref}}[\varphi]\rangle_{\text{ref}}, \tag{34}$$

where the average is taken in the reference ensemble with action $L_{\text{ref}}$. Minimizing the free energy functional with respect to $\phi$ and $G$, we obtain the self-consistent equations [47,52]

$$-\nabla \cdot (\epsilon \nabla \phi) = \lambda_+ z_+ e^{-z_+\phi - u_+} + \lambda_- z_- e^{-z_-\phi - u_-}, \tag{35}$$

$$-\nabla \cdot [\epsilon \nabla G(r,r')] + \left(\lambda_+ z_+^2 e^{-z_+\phi - u_+} + \lambda_- z_-^2 e^{-z_-\phi - u_-}\right) G(r,r') = \delta(r,r'), \tag{36}$$

$$u_\pm(r) = \frac{z_\pm^2}{2} \lim_{r' \to r}\left[ G(r,r') - \frac{1}{4\pi\epsilon |r-r'|}\right]. \tag{37}$$

The term $u_\pm(r)$ appeared in the exponent is called the self-energy that includes the correlation and fluctuation effects.

The continuous FT approach is generalized to a hardcore-ion model by Loubet et al. [49] to study the liquid-vapor phase transformation. A dipolar PB theory [53] is also developed by treating water molecules as explicit dipoles rather than implicit continuous medium.

The FT approach is useful in calculating dielectric constants [54] but the validity of the Gaussian functional is questionable [51]. We have previously shown that the RDFs calculated by the self-consistent equations, Eqs. (35)-(37), match the molecular dynamics (MD) simulation results only for counter-ions [21]. In addition to the variational methods, a systematic perturbative expansion [41] has been performed by Netz and Orland to solve the partition function by a loop-wise expansion of the action around the saddle point, which indicates that the zero-loop equation is just the ordinary PB equation and the one-loop correction yields similar results as the variational methods.

**4. Modified PB equations with correlations**

The ion number distribution in the PB theory is assumed to satisfy the Boltzmann distribution

$$\rho(r) = \rho_0 \exp(-\beta q \psi(r)), \tag{38}$$

where $\rho_0$ is the bulk number density and $q = ze$ is the amount of charge.

Since the Boltzmann distribution is exact only for ideally non-interacting particles, the difference between $q\psi(r)$ and PMF $w(r)$ reflects the physical features neglected in the PB theory, i.e., ion size, molecular details of solvent, fluctuation, and ion correlation. On the other hand, the ion distribution with respect to a reference particle, i.e., the RDF, has the following exact relationship with the PMF (Eq. (8)):

$$g(r) = \rho_0 \exp(-\beta w(r)). \tag{39}$$

Therefore, a more accurate theory may be achieved by dealing with the PMF. In this section, we introduce a correlation-enhanced PB equation developed by us [21], as well as the more sophisticated modified PB theories developed by Outhwaite and Bhuiyan [55-57].

### 4.1. Correlation-Enhanced PB model

Considering a two-species ionic system with valences $z_\pm$ and an average electrostatic potential $\psi(r)$, the correlation effect can be quantified by inserting a test ion. We initially assume that the system obeys the PB equation, so that when the system is fully relaxed to a new equilibrium state after the insertion of the test ion, the average potential is perturbed to be $\psi(r) + G(r, r')$, where $G(r, r')$ is the incremental potential due to the test ion. For simplicity, we define the dimensionless potential $\phi \equiv \beta e \psi$, exactly the same as the one defined in section 3.2, and the Bjerrum length $l_B \equiv \dfrac{\beta e^2}{4\pi \varepsilon_0 \varepsilon}$, then the PB equation with and without the test ion is

$$-\nabla^2 [\phi_{PB}(r) + G_\pm(r, r')] = 4\pi l_B [z_+ \rho_{s+} e^{-z_+ \phi_{PB}(r) - z_+ G_\pm(r,r')} + z_- \rho_{s-} e^{-z_- \phi_{PB}(r) - z_- G_\pm(r,r')} + z_\pm \delta(r - r')] \tag{40}$$

and

$$-\nabla^2 \phi_{PB} = 4\pi l_B \left( z_+ \rho_{s+} e^{-z_+ \phi_{PB}} + z_- \rho_{s-} e^{-z_- \phi_{PB}} \right), \tag{41}$$

where the subscript 'PB' stands for the solution to the original PB equation, $\rho_{s+}$ and $\rho_{s-}$ are bulk number densities of the two ion species. By comparing the two equations, we obtain the expression for $G(r, r')$:

$$-\nabla^2 G_\pm(r, r') = 4\pi l_B [z_+ \rho_{s+} e^{-z_+ \phi_{PB}(r)} \left( e^{-z_+ G_\pm(r,r')} - 1 \right) + z_- \rho_{s-} e^{-z_- \phi_{PB}(r)} \left( e^{-z_- G_\pm(r,r')} - 1 \right) + z_\pm \delta(r - r')]. \tag{42}$$

Eq. (42) indicates that two factors contribute to $G_\pm(r,r')$: the presence of the test ion and the change of surrounding ions due to the presence of the test ion. Therefore, the correlation effect caused by the insertion of the test ion can be quantified by subtracting $G(r,r')$ from the potential of the test ion itself $G_0(r,r')$, and the correlation term $u(r)$, which is just the self-energy, is determined by

$$u_\pm(r') = \lim_{r \to r'} u_\pm(r,r') = \lim_{r \to r'} [G_\pm(r,r') - G_{0\pm}(r,r')]. \tag{43}$$

The self-energy is then included in the original PB equation to obtain the correlation-enhanced PB equation as [21]:

$$-\nabla^2 \phi = 4\pi l_B \left( z_+ \rho_{0+} e^{-z_+\phi - z_+ u_+} + z_- \rho_{0-} e^{-z_-\phi - z_- u_-} \right), \tag{44}$$

where $\rho_{0\pm} = \rho_{s\pm} \exp(z_\pm u_\infty)$ with $u_\infty$ being the self-energy at infinity is the rescaled bulk number density to ensure that local density approaches the bulk value when $r$ goes to infinity. As a modified PB equation is obtained, the potential $\phi_{PB}$ in Eq. (42) should be replaced by the solution of Eq. (44) to obtain the self-consistent equations.

Eqs. (42)-(44) are similar to the self-consistent equations in the FT approach. It can be shown that the formula of our correlation-enhanced PB model is reduced to the FT approach if one applies a linear approximation to the equation for $G(r,r')$ [21]. However, such a linearization is in fact mathematically problematic since $G(r,r')$ diverges as $r'$ goes to $r$, and the linearization is acceptable only when $G(r,r')$ is close to zero, indicating that the widely used Gaussian reference action is *ad hoc* and may not be a good choice for charged particles systems. The RDFs obtained by the MD simulations for 1:1 and 1:2 ionic solutions show that our correlation-enhanced PB model matches the MD results very well but the corresponding FT equation deviate significantly from the MD results [21].

**4.2. Outhwaite-Bhuiyan modified PB (MPB) theories**

Outhwaite and Bhuiyan developed a series of MPB theories known as MPB1 to MPB5 yielding different levels of accuracy [55,56] with MPB5 the most accurate one. The MPB theories can be considered as a generalization of the integral equation method (see Appendix) to the heterogeneous case. The equation describing the model composed of a charged hard-sphere particle next to a planar hard electrode with a uniform surface charge density $\sigma$ is [58]

$$\frac{d^2\psi}{dx^2} = -\frac{1}{\varepsilon_0\varepsilon}\sum_s q_s\rho_s g_s(x), \tag{45}$$

where $g_s(x)$ is the singlet wall-ion distribution function which can be obtained using the Kirkwood charging process [59,60] by

$$g_s(x) = \xi_s(x)\exp\left[-\beta q_s\psi(x) - \beta\int_0^{e_s}\lim_{r\to 0}(\varphi_s^* - \varphi_{sb}^*)dq_s\right], \tag{46}$$

where $\xi_s(x) = g_s(x|q_s=0)$ is the exclusive volume term, $\varphi^*$ is the fluctuation potential, the subscript $b$ denotes the corresponding value for bulk, and the subscript $s$ denotes the ion species. After complicated calculation of the fluctuation potential, the expression finally appears to be

$$g_s(x) = \xi_s(x)\exp\left[-\beta q_s L(\psi) - \frac{\beta q^2}{8\pi\varepsilon_0\varepsilon}(F - F_0)\right], \tag{47}$$

where the operator $L(\psi)$ is given by

$$L(\psi) = \frac{F}{2}\{\psi(x+a) + \psi(x-a)\} - \frac{(F-F_0)}{2a}\int_{x-a}^{x+a}\psi(X)dX, \tag{48}$$

and the coefficients $F$ and $F_0$ are expressed in terms of the distance $x$ and ion diameter $a$. Details of the derivation can be found in Ref. [56]. The expression for the exclusive volume term can be obtained by applying the Born-Green-Yvon hierarchy [61], or by using an alternative and simpler expression developed by Outhwaite and Lamperski [62]:

$$\xi_s(x) = \xi_s^0(x)\exp\left[\int\sum_t \rho_t c_{st}^0\{g_t(x) - g_t(x|\sigma=0)\}dV\right], \tag{49}$$

where $c_{st}^0$ is the inhomogeneous neutral-sphere direct correlation function whose meaning can be found in the Appendix and $\xi_s^0(x) = \xi_s(x|\sigma=0)$. The $c_{st}^0$ and $\xi_s^0(x)$ can be approximated by the PY results [63,64] and the subscript $t$ runs over all the ion species.

The MPB equations are solved for a primitive model of a planar double layer containing 1:1 or 1:2 charged hard-sphere particles. The results are compared with corresponding Monte Carlo (MC) simulation results and both the MPB formulations reproduce the MC data very well [58].

**5. Mean field theories**

The modified PB equations described above are usually hard or impossible to be solved analytically, which limits their practical applications in some scenarios. A simple analytic expression for the ion or potential distribution is important and convenient in many applications. In this section, we introduce two mean-field theories, the DIT [37,38,65] and the MDH theory [39,40]. The former provides an effective-charge Yukawa potential and the latter ends up with a multi-Yukawa potential.

### 5.1. The dressed-ion theory (DIT)

To derive the DIT, we start with the Ornstein-Zernike (OZ) equation for the pair correlation functions:

$$h_{ij}(r) = c_{ij}(r) + \sum_{l} \int dr' \, c_{il}(|r-r'|) n_l h_{lj}(r'), \tag{50}$$

where $h(r)$ is the total correlation function (TCF) and $c(r)$ is the direct correlation function (DCF), see the Appendix for their descriptions. It is convenient to apply the Fourier transform to obtain

$$\hat{h}_{ij}(k) = \hat{c}_{ij}(k) + \sum_{l} \hat{c}_{il}(k) n_l \hat{h}_{lj}(k) \tag{51}$$

and convert it to the matrix form

$$\hat{\boldsymbol{H}} = \hat{\boldsymbol{C}} + \hat{\boldsymbol{C}} \boldsymbol{N} \hat{\boldsymbol{H}}, \tag{52}$$

where $\hat{\boldsymbol{H}} = \{\hat{h}_{ij}\}$, $\hat{\boldsymbol{C}} = \{\hat{c}_{ij}\}$, $\boldsymbol{N} = \{\delta_{ij} n_i\}$. The total charge distribution is

$$\rho_j^{\text{tot}}(r) = q_j \delta^{(3)}(r) + \sum_{l} q_l n_l h_{lj}(r) \tag{53}$$

and the average potential is

$$\psi_j(r) = \int dr' \, \varphi(|r-r'|) \rho_j^{\text{tot}}(r'), \tag{54}$$

where $\varphi(r) = 1/(4\pi\varepsilon\varepsilon_0 r)$ is the Coulomb potential. In the Fourier space the average potential is

$$\hat{\psi}_j(k) = \hat{\varphi}(k) \hat{\rho}_j^{\text{tot}}(k) = \hat{\varphi}(k) \left[ q_j + \sum_{l} q_l n_l \hat{h}_{lj}(k) \right] \tag{55}$$

and in the matrix form

$$(\hat{\boldsymbol{\psi}})^{\text{T}} = \hat{\varphi} \left[ \boldsymbol{q}^{\text{T}} + \boldsymbol{q}^{\text{T}} \boldsymbol{N} \hat{\boldsymbol{H}} \right], \tag{56}$$

where $\hat{\boldsymbol{\psi}} = \{\hat{\psi}_j\}$, $\boldsymbol{q} = \{q_j\}$ are vectors.

The key idea of the DIT is to split the pair correlation function into a short-range part and a long-range part, then the DCF is

$$c_{ij}(r) = c_{ij}^0(r) + c_{ij}^l(r), \tag{57}$$

where superscript 0 denotes the short-range part and superscript $l$ denotes the long-range part. Analogous to the long-range behavior of the RDF, the long-range part of DCF $c_{ij}^l(r) \sim -\beta q_i q_j \varphi(r)$ when $r \to \infty$. Therefore, we have

$$c_{ij}^0(r) = c_{ij}(r) + bq_i q_j j(r), \tag{58}$$

and the Fourier transform in the matrix form is

$$\hat{\boldsymbol{C}}^0 = \hat{\boldsymbol{C}} + \beta \boldsymbol{q}\boldsymbol{q}^{\mathrm{T}} \hat{\varphi}. \tag{59}$$

Inserting the short-range part of DCF into the OZ equation and utilizing Eq. (56), we have

$$\hat{\boldsymbol{H}} = \hat{\boldsymbol{C}}^0 + \hat{\boldsymbol{C}}^0 \boldsymbol{N} \hat{\boldsymbol{H}} - \beta \boldsymbol{q}(\hat{\boldsymbol{\psi}})^{\mathrm{T}}. \tag{60}$$

The solution of the TCF is

$$\hat{\boldsymbol{H}} = \left(\boldsymbol{1} - \hat{\boldsymbol{C}}^0 \boldsymbol{N}\right)^{-1} \hat{\boldsymbol{C}}^0 - \beta \left(\boldsymbol{1} - \hat{\boldsymbol{C}}^0 \boldsymbol{N}\right)^{-1} \boldsymbol{q}(\hat{\boldsymbol{\psi}})^{\mathrm{T}}. \tag{61}$$

Defining the short-range variables $\hat{\boldsymbol{H}}^0 = \left(\boldsymbol{1} - \hat{\boldsymbol{C}}^0 \boldsymbol{N}\right)^{-1} \hat{\boldsymbol{C}}^0$ and $\hat{\boldsymbol{\rho}}^0 = \left(\boldsymbol{1} - \hat{\boldsymbol{C}}^0 \boldsymbol{N}\right)^{-1} \boldsymbol{q}$, we have

$$\hat{\boldsymbol{H}} = \hat{\boldsymbol{H}}^0 - \beta \hat{\boldsymbol{\rho}}^0 (\hat{\boldsymbol{\psi}})^{\mathrm{T}}. \tag{62}$$

In the real space, the TCF is

$$h_{ij}(r) = h_{ij}^0(r) - \beta \int d\boldsymbol{r}' \rho_i^0(|\boldsymbol{r} - \boldsymbol{r}'|) \psi_j(\boldsymbol{r}'). \tag{63}$$

The form of Eq. (63) is the same as that of the DH theory, but their definitions of the charge densities are different. In the DH theory, the corresponding expression for the TCF is

$$h_{ij}^{\mathrm{DH}}(r) = -\beta \int d\boldsymbol{r}' q_i \delta^{(3)}(r) \psi_j^{\mathrm{DH}}(\boldsymbol{r}'). \tag{64}$$

where $\delta^{(3)}(r)$ is the three-dimensional Dirac delta function. For DIT, to derive an expression for the average potential, we first express $\rho^0$ in terms of $h_{ij}$:

$$\hat{\boldsymbol{\rho}}^0 = \boldsymbol{q} + \hat{\boldsymbol{H}}^0 \boldsymbol{N} \boldsymbol{q}. \tag{65}$$

In the real space

$$\rho_j^0(r) = q_j \delta^{(3)}(r) + \sum_l q_l n_l h_{lj}^0(r), \tag{66}$$

Combining Eqs. (53), (63), and (66), we obtain the total density

$$\begin{aligned}\rho_j^{tot}(r) &= q_j \delta^{(3)}(r) + \sum_l q_l n_l h_{lj}(r) \\ &= \rho_j^0(r) - \beta \int dr' \sum_l q_l n_l \rho_l^0(|r-r'|) \psi_j(r') \\ &= \rho_j^0(r) - \int dr' \alpha(|r-r'|) \psi_j(r'),\end{aligned} \tag{67}$$

where $\alpha(|r-r'|) \equiv \sum_l q_l n_l \rho_l^0(|r-r'|)$ is a response function comparable to the Debye parameter $\kappa_{DH}^2 = \beta \sum_l n_l q_l^2$.

The average potential satisfies Poisson's equation:

$$-\varepsilon_0 \varepsilon \nabla^2 \psi_j(r) = \rho_j^{tot}(r) = \rho_j^0(r) - \int dr' \alpha(|r-r'|) \psi_j(r'), \tag{68}$$

which reduces to the DH equation if we replace $\rho_j^0(r)$ with the ion density and $\alpha(|r-r'|)$ with $\kappa_{DH}^2$. We see that all information we need to obtain the solution is the expression of $\alpha(|r-r'|)$. Applying the linear response theory [66], it can be shown that $\alpha(r)$ is related to the dielectric function $\hat{\varepsilon}_l(k)$ in the Fourier space by

$$\hat{\varepsilon}(k) = \varepsilon + \frac{\hat{\alpha}(k)}{\varepsilon_0 k^2}, \tag{69}$$

where $\hat{\alpha}(k)$ is the Fourier transform of $\alpha(r)$. Therefore, all the information of the key parameters in DIT is contained in the response functions. Practically, the linear response function $\alpha(r)$ can be obtained by experiment and/or MD simulation. In the theoretical work by Varela et al. [67], $\alpha(r)$ is also obtained by calculating the static structure factor for a one-component charged-sphere model in the modified mean-spherical approximation (MSA) and the random phase approximation (RPA). The authors have shown that the RPA clearly accounts for the basic feature of the DIT that the distribution functions split into two well-defined parts and can be considered as a first-order approximation to the problem [12].

### 5.2. The molecular Debye-Hückel (MDH) theory

The MDH theory, which is closely related to the DIT, still needs the information of the dielectric function to determine its key factors. An advantage of the MDH theory is that the coefficients are not too sensitive to the size and charge of the solute molecules [39], which allows the usage of the solvent to estimate the coefficients for the solute. The Gaussian unit is adopted and the ^ on top of the variables in the Fourier space is omitted in the following equations, as is in the original paper.

To derive the MDH theory, we start with the Poisson's equation in the Fourier space ([40]) of the DIT:

$$k^2 \varepsilon(k) \psi_j(k) = 4\pi \rho_j^0(k), \tag{70}$$

in which $\varepsilon(k) = \varepsilon + \dfrac{4\pi \alpha(k)}{k^2}$ is the dielectric function of the solution with the relative dielectric constant $\varepsilon$ and $\alpha(k)$ is just $\hat{\alpha}(k)$ in Eq. (69). The charge density $\rho_j^0(k)$ has the same definition as in Eq. (66), but it can be interpreted as the free charge density in most cases as long as the linear response theory is applicable [40].

The roots of $\varepsilon(k)$ are useful for finding the solution and are denoted as $ik_n$, therefore

$$k^2 \varepsilon(k) = \dfrac{\varepsilon \prod_n (k^2 + k_n^2)}{f(k)}, \tag{71}$$

where $f(k)$ is a function with no poles. We can then solve the potential in the real space using the inverse Fourier transform

$$\begin{aligned}\psi_j(r) &= \dfrac{1}{2\pi^2 r} \int_0^\infty dk \left[ \psi_j(k) k \right] \sin(kr) \\ &= \dfrac{1}{2\pi^2 r} \int_0^\infty dk \left[ \dfrac{k f(k) \rho_j^0(k)}{\varepsilon \prod_n (k^2 + k_n^2)} \right] \sin(kr). \end{aligned} \tag{72}$$

Solving the integral using the residue theorem, we have the following asymptotic solution:

$$\psi_j(r) \sim \sum_n \dfrac{q_{\text{eff},jn}}{\varepsilon_{\text{eff},n}} \dfrac{e^{-k_n r}}{r}, \tag{73}$$

where $q_{\text{eff},jn} = \rho_j^0(ik_n)$ is an effective charge and $\varepsilon_{\text{eff},n} = \dfrac{1}{2}\left[ k \dfrac{d\varepsilon(k)}{dk} \right]_{k=ik_n}$ is an effective dielectric constant. The solution is a linear combination of multiple Yukawa potentials with different Debye lengths $k_n$. The DH theory can be obtained by applying $\varepsilon(k) = \varepsilon(1 + k_{\text{DH}}^2/k^2)$ to the MDH theory so that it has only one root $ik_{DH}$ and hence results in a Yukawa potential.

Since the solution of the MDH equation is a combination of the solutions of DH equations with various Debye lengths, it can be determined by finding an effective linearly-combined coefficient set instead of using the effective parameters directly. Given the dielectric function is known, this kind of coefficient set can be determined by the Stillinger-Lovett relation [68-70]. In practice, the dielectric function can be calculated by the MSA or the HNC approximation.

6. Conclusions

In this paper, we review the original PB theory and some recent progresses of improving it. The original PB theory is the first successful theoretical model for describing charged systems including ionic solutions, but the neglect of ion correlations and exclusive volume effects disables its applicability to the cases of high charge densities or valences.

The FT approach, aided by the definition of self-energy, provides a way to directly solve the grand partition function by performing the HS transformation, and various approximation methods lead to different versions of the FT approach. The variational method with a Gaussian reference action was used by several groups to solve the functional form of the partition function. However, the validity of the Gaussian functional is questionable, and we found its numerical solution deviates significantly from MD simulation results.

A correlation-enhanced PB model was developed by us based on physical pictures, which also defines a self-energy but whose expression is different from the FT approach. We have shown that our model reduces to the variational method of the FT approach when applying a linear approximation, which is nevertheless mathematically problematic. A more sophisticated modified PB theory was developed by Outhwaite and Bhuiyan, which still keeps the form of the original PB equation, but the pair correlation functions are calculated analytically with the integral equation theory.

The mean field theories try to find a solution as simple as the Yukawa potential. The DIT gives an exact expression, which has the form of the DH theory with the variables to be determined by splitting the pair correlation function into a short-range part and a long-range part. The MDH theory suggests a multi-Yukawa potential which is also exact as long as the coefficients are precisely determined. However, both the DIT and MDH theories require the information of dielectric function, which needs to be further determined either by theoretical approximations or experimental data.

**Appendix: Integral equation theory**

The direct correlation function $c(r)$ and total correlation function $h(r)$ are related by the OZ equation:

$$h_{ij}(r) = c_{ij}(r) + \sum_l \int dr' c_{il}(|r-r'|) n_l h_{lj}(r'). \tag{A1}$$

The OZ equation can be understood as follows: the particles $i$ and $j$ can be related directly by $c_{ij}(r)$, or indirectly through another particle $l$. The total correlation function is defined as

$$h(r) = g(r) - 1, \tag{A2}$$

where $g(r)$ is the RDF. Once we obtain the total correlation function or equivalently the RDF, the properties of the system can be calculated with the knowledge of statistical physics. However, the direct correlation function is hard to determine, not to mention that its physical meaning is not well defined. To solve the OZ equation, several approximate methods, known as the integral equation theories [24], are proposed.

An intuitive method is the MSA, which is defined in terms of the direct correlation function by:

$$\begin{aligned} g(r) &= 0, & r < d \\ c(r) &= -\beta v(r), & r > d \end{aligned} \tag{A3}$$

where $d$ is the hard sphere diameter and $v(r)$ is the interaction potential between particles. In the MSA, the OZ equation can be solved analytically and can be shown to reduce to the DH theory at a high temperature or a low density limit.

Other classic approximations are the PY and HNC equations:

$$\begin{aligned} g(r) &= \exp[-\beta v(r)][1 + h(r) - c(r)] & \text{(PY)}, \\ g(r) &= \exp[-\beta v(r)] \exp[h(r) - c(r)] & \text{(HNC)}. \end{aligned} \tag{A4}$$

The PY or HNC equation can be solved numerically together with the OZ equation. The HNC approximation can reproduce simulation results for charged particles systems very accurately, while the PY approximation is proved to be more successful for short-range potentials.

**Corresponding Author**

*E-mail: wangyt@itp.ac.cn. Phone: +86 10-62648749.

**Acknowledgement**

This work was supported by the Strategic Priority Research Program of Chinese Academy of Sciences (Grant No. XDA17010504), the National Natural Science Foundation of China (Nos.

**References**

[1] R. Wang and Z. G. Wang, J. Chem. Phys. **139**, 124702 (2013).

[2] R. Wang and Z. G. Wang, Phys. Rev. Lett. **112**, 136101 (2014).

[3] R. M. Adar, D. Andelman, and H. Diamant, Phys. Rev. E **94**, 9, 022803 (2016).

[4] Z. Tan, W. Zhang, Y. Shi, and F. Wang, in *Advance in Structural Bioinformatics*, edited by D. Wei *et al.* (Springer, Dordrecht, 2015), pp. 143.

[5] Z.-J. Tan and S.-J. Chen, J. Chem. Phys. **122**, 044903 (2005).

[6] Z. J. Tan and S. J. Chen, Biophys. J. **90**, 1175 (2006).

[7] P. Debye and E. Huckel, Physikalische Zeitschrift **24**, 185 (1923).

[8] D. Andelman, in *Handbook of Biological Physics* (Elsevier, Amsterdam, 1995), pp. 603.

[9] Z. Mester and A. Z. Panagiotopoulos, J. Chem. Phys. **142**, 044507 (2015).

[10] Y. Levin, Rep. Prog. Phys. **65**, 1577 (2002).

[11] B. Honig and A. Nicholls, Science **268**, 1144 (1995).

[12] L. M. Varela, M. Garcia, and V. Mosquera, Phys. Rep. **382**, 1 (2003).

[13] P. Grochowski and J. Trylska, Biopolymers **89**, 93 (2008).

[14] R. Messina, J. Phys.-Condes. Matter **21**, 113102, 113102 (2009).

[15] H. Li and B. Lu, J. Chem. Phys. **141**, 024115 (2014).

[16] X. Liu, Y. Qiao, and B. Lu, SIAM J. Appl. Math. **78**, 1131 (2018).

[17] D. A. McQuarrie, *Statistical Mechanics* (Harper, New York, 1976).

[18] D. Chandler, *Introduction to Modern Statistical Mechanics* (Oxford University Press, London, 1987).

[19] G. H. Bolt, J. Colloid Sci. **10**, 206 (1955).

[20] B. Z. Lu, Y. C. Zhou, M. J. Holst, and J. A. McCammon, Commun. Comput. Phys. **3**, 973 (2008).

[21] M. Su, Z. Xu, and Y. Wang, J. Phys.-Condes. Matter **31**, 355101 (2019).

[22] G. Ren and Y. Wang, Eur. Phys. Lett. **107**, 30005 (2014).

[23] B. t. Roux and T. Simonson, Biophys. Chem. **78**, 1 (1999).

[24] J.-P. Hansen and I. R. McDonald, in *Theory of Simple Liquids (Third Edition)*, edited by J.-P. Hansen, and I. R. McDonald (Academic Press, Burlington, 2006), pp. 78.

[25] A. Y. Grosberg, T. T. Nguyen, and B. I. Shklovskii, Rev. Mod. Phys. **74**, 329 (2002).

[26] B. I. Shklovskii, Phys. Rev. E **60**, 5802 (1999).

[27] E. Gonzalez-Tovar, F. Jimenez-Angeles, R. Messina, and M. Lozada-Cassou, J. Chem. Phys. **120**, 9782 (2004).

[28] R. Messina, C. Holm, and K. Kremer, Phys. Rev. E **64**, 021405 (2001).

[29] J. Yu, G. E. Aguilar-Pineda, A. Antillon, S. H. Dong, and M. Lozada-Cassou, J. Colloid Interface Sci. **295**, 124 (2006).

[30] A. Diehl, H. A. Carmona, and Y. Levin, Phys. Rev. E **64**, 6, 011804 (2001).

[31] N. Gronbech-Jensen, R. J. Mashl, R. F. Bruinsma, and W. M. Gelbart, Phys. Rev. Lett. **78**, 2477 (1997).

[32] B. Y. Ha and A. J. Liu, Phys. Rev. Lett. **79**, 1289 (1997).

[33] M. Kardar and R. Golestanian, Rev. Mod. Phys. **71**, 1233 (1999).

[34] R. Podgornik and V. A. Parsegian, Phys. Rev. Lett. **80**, 1560 (1998).

[35] C. Lin, X. Zhang, X. Qiang, J.-S. Zhang, and Z.-J. Tan, J. Chem. Phys. **151**, 114902 (2019).

[36] L. Onsager, Chem. Rev. **13**, 73 (1933).

[37] R. Kjellander, *Distribution function theory of electrolytes and electrical double layers - Charge renormalisation and dressed ion theory* (Springer, Dordrecht, 2001), Vol. 46, Electrostatic Effects in Soft Matter and Biophysics.

[38] R. Kjellander and D. J. Mitchell, J. Chem. Phys. **101**, 603 (1994).

[39] T. Xiao, ChemPhysChem **16**, 833 (2015).

[40] T. Xiao and X. Song, J. Chem. Phys. **135**, 104104, 104104 (2011).

[41] D. Frydel, Eur. J. Phys. **36**, 18, 065050 (2015).

[42] R. R. Netz and H. Orland, Europhys. Lett. **45**, 726 (1999).

[43] A. Naji, M. Kanduc, J. Forsman, and R. Podgornik, J. Chem. Phys. **139**, 13 (2013).

[44] J. Hubbard, Phys. Rev. Lett. **3**, 77 (1959).

[45] R. R. Netz and H. Orland, Eur. Phys. J. E **1**, 203 (2000).

[46] R. R. Netz and H. Orland, Eur. Phys. J. E **11**, 301 (2003).

[47] Z. G. Wang, Phys. Rev. E **81**, 021501, 021501 (2010).

[48] Z. L. Xu and A. C. Maggs, J. Comput. Phys. **275**, 310 (2014).

[49] B. Loubet, M. Manghi, and J. Palmeri, J. Chem. Phys. **145**, 044107, 044107 (2016).

[50] S. Buyukdagli, M. Manghi, and J. Palmeri, Phys. Rev. E **81**, 041601, 041601 (2010).

[51] S. Buyukdagli and R. Blossey, J. Phys.-Condes. Matter **28**, 343001, 343001 (2016).

[52] R. Wang and Z. G. Wang, J. Chem. Phys. **142**, 104705 (2015).

[53] A. Abrashkin, D. Andelman, and H. Orland, Phys. Rev. Lett. **99**, 077801 (2007).

[54] A. Levy, D. Andelman, and H. Orland, Phys. Rev. Lett. **108**, 227801 (2012).

[55] C. W. Outhwaite, L. B. Bhuiyan, and S. Levine, J. Chem. Soc., Faraday Trans. II **76**, 1388 (1980).

[56] C. W. Outhwaite and L. B. Bhuiyan, J. Chem. Soc., Faraday Trans. II **79**, 707 (1983).

[57] C. W. Outhwaite, M. Molero, and L. B. Bhuiyan, J. Chem. Soc., Faraday Trans. **87**, 3227 (1991).

[58] L. Bari Bhuiyan and C. W. Outhwaite, Phys. Chem. Chem. Phys. **6**, 3467 (2004).

[59] J. G. Kirkwood, J. Chem. Phys. **2**, 767 (1934).

[60] J. G. Kirkwood and J. C. Poirier, J. Phys. Chem. **58**, 591 (1954).

[61] C. W. Outhwaite and L. B. Bhuiyan, J. Chem. Soc., Faraday Trans. II **78**, 775 (1982).

[62] C. W. Outhwaite and S. Lamperski, Condens. Matter Phys. **4**, 739 (2001).

[63] S. Lamperski and L. B. Bhuiyan, J. Electroanal. Chem. **540**, 79 (2003).

[64] S. Lamperski and C. W. Outhwaite, Langmuir **18**, 3423 (2002).

[65] R. Kjellander and D. J. Mitchell, Chem. Phys. Lett. **200**, 76 (1992).

[66] J.-P. Hansen and I. R. McDonald, in *Theory of Simple Liquids (Third Edition)*, edited by J.-P. Hansen, and I. R. McDonald (Academic Press, Burlington, 2006), pp. 291.

[67] L. M. Varela, M. Perez-Rodriguez, M. Garcia, F. Sarmiento, and V. Mosquera, J. Chem. Phys. **109**, 1930 (1998).

[68] R. Lovett and F. H. Stillinger, J. Chem. Phys. **48**, 3869 (1968).

[69] F. H. Stillinger and R. Lovett, J. Chem. Phys. **49**, 1991 (1968).

[70] F. H. Stillinger and R. Lovett, J. Chem. Phys. **48**, 3858 (1968).